\documentclass[twocolumn, secnumarabic,  floatfix, nofootinbib, nobalancelastpage]{revtex4-2}

\def\TitleOfPaper{Using Machine Learning to Find New Density Functionals}
\def\preprintlinklocation{http://dft.uci.edu + PAPER REF}

\usepackage{amsmath}
\usepackage{physics}
\usepackage{amssymb}
\usepackage{float}
\usepackage[dvipsnames]{xcolor}

\definecolor{TITLECOL}{rgb}{0.05,0.25,0.85}
\definecolor{CONTENTSCOL}{rgb}{0.1,0.2,0.7}
\definecolor{URLCOL}{rgb}{0,0.52,0.83}
\definecolor{LINKCOL}{rgb}{0.05,0.5,0}
\definecolor{CITECOL}{rgb}{0.25,0,0.48}
\definecolor{SECOL}{rgb}{0,0,0}
\definecolor{SSECOL}{rgb}{0.26,0.19,0.75}

\renewcommand{\sec}[1]{\section{\textcolor{SECOL}{#1}}}

\usepackage{graphicx}

\def\PublicationsLink{http://dft.uci.edu/publications.php}
\def\preprintlink{ \href{\preprintlinklocation}{\TitleOfPaper} }
\def\preprinttext{~}

\usepackage{fancyhdr}
\makeatletter
\fancypagestyle{titlepage}
{	\lhead{\textsc{\preprinttext\ ~ \href{\PublicationsLink}{~}}}
	\chead{}
	\rhead{\preprintlink}
	\lfoot{}}
\makeatother
\def\preprintlink{ 
	\href{\preprintlinklocation}
        {%v\versionnumber\ - 
~}%INSERT the reference of your paper
	}

\definecolor{Green}{rgb}{0.016,0.627,0}
\definecolor{Plum}{rgb}{0.17,0,0.45}
\definecolor{LBlue}{rgb}{0,0.34,0.45}
\definecolor{Sepia}{rgb}{0.37,0.17,0.02}
\definecolor{BurntOrange}{rgb}{0.78,0.39,0}

\usepackage{hyperref}
\hypersetup{
    breaklinks=true,
    bookmarksopen=true,
    bookmarksnumbered=true,
    colorlinks=true,
    linkcolor=LINKCOL,
    linktocpage=true,
    citecolor=CITECOL,
    filecolor=magenta,      
    urlcolor=URLCOL,
}

\def\bea{\begin{eqnarray}}
\def\eea{\end{eqnarray}}
\def\ben{\begin{equation}}
\def\een{\end{equation}}
\def\benu{\begin{enumerate}}
\def\enu{\end{enumerate}}

\def\bei{\begin{itemize}}
\def\eei{\end{itemize}}
\def\beit{\begin{itemize}}
\def\eit{\end{itemize}}
\def\benu{\begin{enumerate}}
\def\enu{\end{enumerate}}

\def\sss{\scriptscriptstyle\rm}

\def\1var{(\bx_1...\bx\N)}

\def\bx{{x}}

\def\N{_{\sss N}}

\def\sph_int{ {\int d^3 r}}

\graphicspath{{figures/}}

\begin{document}
\sf
\title{\TitleOfPaper}

%Put the names of authors here:
\author{Bhupalee Kalita}
\email{bkalita@uci.edu}
\affiliation{Departments of Chemistry, 
University of California, Irvine}

\author{Kieron Burke}
\email{kieron@uci.edu}
\affiliation{Departments of Chemistry, 
University of California, Irvine}
\affiliation{Departments of Physics and Astronomy, 
University of California, Irvine}

\begin{abstract}
    Machine learning has now become an integral part of research and innovation. The field of machine learning density functional theory has continuously expanded over the years while making several noticeable advances. We briefly discuss the status of this field and point out some current and future challenges. We also talk about how state-of-the-art science and technology tools can help overcome these challenges. This draft is a part of the "Roadmap on Machine Learning in Electronic Structure" to be published in \textit{Electronic Structure (EST)}.
\end{abstract}
\maketitle
%\tableofcontents

\sec{Status}

Density functional theory (DFT) has provided low-cost alternatives to direct solution of the Schrödinger equation for almost a century~\cite{T27}.  The Kohn-Sham (KS) scheme~\cite{KS65}, in which only the exchange-correlation (XC) energy needs to be approximated as a functional of the density, has greatly improved accuracy while maintaining low computational cost. Today, about 30\% of supercomputer use is devoted to solving these equations, but there are hundreds of different human-designed XC approximations in use, each producing different predictions.  Almost all begin using the density and its gradient (semilocal approximations). Materials science is dominated by simple standard functionals, often designed using exact conditions, while chemistry mostly uses approximations designed only for molecules, but often achieving higher accuracy.

Four prominent limitations come to mind. Most DFT calculations are for weakly correlated systems, and there is tremendous desire to improve their accuracy without significant computational cost. Second, DFT has well-known generic failures, such as self-interaction error or poor energetics for strongly correlated systems, such as a stretched H$_2$ molecule~\cite{PRSNK21}. Most XC approximation fail to produce a realistic binding energy curve without breaking spin. As one goes from two atoms to four and many, the difficulties grow and can be related to the failure of DFT approximations to capture Mott-Hubbard physics~\cite{SWWB12}. Third, theorems prove that, in principle, one can avoid solving the KS equations if one has a sufficiently accurate approximation for the KS kinetic energy, but here the limitations are even greater, due to the need to extract accurate densities and total energies. Finally, ground-state DFT yields only ground-state energies and densities, but there is also tremendous need to predict response properties.  Here we focus only on the ground state.

Machine learning (ML) has already helped with functional development. In prescient work a quarter-century ago, Tozer et al~\cite{TIH96} found a semilocal approximation by training a neural network to optimize a fit to KS potentials. Moreover, Bayesian methods were used to analyse DFT errors in 2005~\cite{MKFJ05}. More recently, Snyder et al~\cite{SRHB12} used kernel ridge regression (KRR) combined with a principal component analysis of training densities, to create a KS kinetic energy functional reaching chemical accuracy, albeit in a very simple model. And Nagai et al~\cite{NAS20} showed that, by training a neural network (NN) on both the densities and energies of just a few molecules, one could create semilocal approximations comparable in accuracy to those of humans and generalizing to a broad range of molecules.

The field of using ML to design functionals is in its infancy, and improvements in speed, accuracy, and applicability of DFT are beckoning.  Any such improvements that can be implemented in standard codes will have enormous scientific impact.

\sec{Current and future challenges}

ML is promising for improving density functional approximations to overcome the limitations listed above, and progress is likely in all three areas.

First, there are many ingredients already in use for making XC approximations, including dispersion corrections, fractions of exact exchange (both global and range separated), random phase approximation, etc. Can ML be used to find the ‘best’ combination of these ingredients?  More fundamentally, how do we define ‘best’? 

Second, ML allows the possibility of constructing completely non-local functionals, using information about the density at every point in space, either with KRR or NN’s.  This can be used to find the exact functional for strongly correlated systems, as in Ref.~\cite{LHPB20}, especially if full differentiable programming techniques are used. Here, by using the KS equations as a regularizer, a full dissociation curve for (one-dimensional) H$_2$ was constructed from just two data points alone, suggesting tremendous potential for generalizability. However, such a functional, defined on the whole R-space, cannot be applied to arbitrary systems, so what features must be included to make it work more generally?

Third, ML can produce pure density functionals, which could bypass the need to solve the KS equations. This was demonstrated for small molecules, producing an ML functional that yields accurate densities and energies for malonaldehyde and resorcinol MD simulations~\cite{BVLM17}, and for water in condensed phase~\cite{DF-S20}. But, as above, such functionals cannot be expected to generalize well, and so must be retrained for each new species, unlike standard DFT.

\begin{figure}[htp!]
\includegraphics[width=\columnwidth]{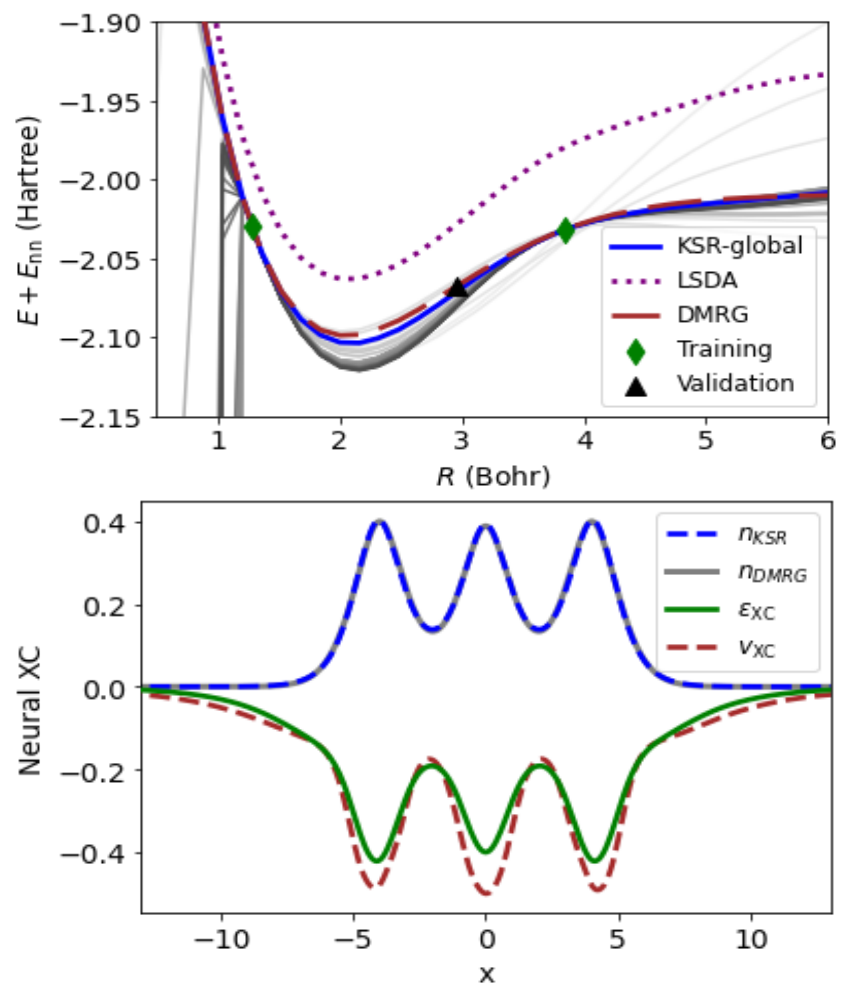}
\caption{Top: One dimensional H$_3$ dissociation energy curve created with the KSR-global function from Ref~\cite{LHPB20}. This model was trained with just two configurations. The changes in predictions as the model evolves from underfitted to overfitted are shown by the darkening shades of grey. The optimal parameters, determined from a single validation configuration, yield the chemically accurate blue curve. $E_{nn}$ is the nuclear interaction energy, DMRG (density matrix renormalization group) is essentially exact, and LSDA is the result of the local spin density approximation. Bottom: The density, XC energy and XC potential of H3 at 4 Bohr, calculated using the optimal parameters. }
\label{fig:h3}
\end{figure}

In Figure.~\ref{fig:h3}, we employ the KS regularizer (KSR) from Ref.~\cite{LHPB20} to calculate the binding curve of 1D H$_3$ and show its attributes at R= 4 Bohr. The KSR is chemically accurate even when the bond is stretched and predicts the density with negligible error. A recent study~\cite{KV21} provides an example of implementation of differentiable DFT in 3D. A similar extension of the work in Ref. [9] can effectively provide a stable solution for strongly correlated matter. However, much work remains to test these algorithms to answer what would be the degree of generalization and what could be done to improve them further.

\sec{Advances in science and technology to meet the challenges}

ML has revolutionized many aspects of everyday life, from movie selection to facial recognition.  Over the past 10-15 years, there have been significant attempts to use it in physical sciences and especially in electronic structure theory.  The most notable success has been the development of force fields, both in chemical and configuration space~\cite{LWMZ21}.

But the development of density functional approximations is still a black art, requiring an unholy alliance of physical (or chemical) intuition, deep knowledge of theory, and some very carefully chosen data. A major difficulty is to build ML models that respect all the implicit (and explicit) rules in 
DFT that humans know (often only intuitively) so that the models extrapolate appropriately to new materials and new molecules.  With our traditional XC approximations, when we run a KS calculation on an entirely new problem, we have a strong sense of how accurate we expect it to be, and certain intuitive consistency tests, such as trying a different functional, even if we cannot put quantitative error bars on our predictions. If we can use ML to design better functionals, overcoming any of the three challenges mentioned previously, such ML-designed functionals will permanently alter the computational landscape.

Much has been said and written about the potential for quantum computers to transform electronic structure calculations. It is certainly true that, once a sufficiently large error-correcting machine is widely available, there are several strongly correlated problems that they might solve for us. But unless there are extreme speedups in routine classical computations, DFT will long continue as the workhorse for the 99\% of problems (or aspects of these problems) that DFT works well for.

\sec{Concluding remarks}
The applications of ML to functional design are still in their infancy.  There is no general-purpose XC approximation designed by ML in use or available in most codes.  It will take more effort and research to understand what the best way is to apply ML techniques (likely NN’s) to develop better approximations, including ones that can be systematically improved with increases in training data.  ML could produce either faster or more accurate functionals for present applications or extend the reach of practical DFT calculations to encompass strongly correlated systems. The future looks bright but has not arrived yet. 

\begin{acknowledgments}
K. B. acknowledges NSF grant no. CHE 1856165 and B.K acknowledges NSF grant no. DGE 1633631.
\end{acknowledgments}

%If someone wants to take the burkebibstyle and put it into rev4-2 instead of rev4-1 that would be great. Use a diff tool for Burke to rev4-1 and then add those in where they need to go in rev4-2.
%\bibliographystyle{lucas-preprint}
\bibliographystyle{apsrev4-2}
\bibliography{Master}

\label{page:end}
\end{document}